\documentclass[aps,prb,preprint,showpacs,preprintnumbers,amsmath,amssymb]{revtex4}
\usepackage{amsmath}
\usepackage{amssymb}
\usepackage{amsfonts}
\usepackage{color,array}
\usepackage{dsfont}
\usepackage{slashed}
\usepackage{graphicx}
\usepackage{epstopdf}
\usepackage{graphics}
\usepackage{epsfig}
\usepackage{bbm,bm}
\usepackage{psfrag}
\usepackage{ulem}
\usepackage{url}

\begin{document}

\title{Anisotropy of Cubic Ferromagnet at Criticality}

\author{A. Kudlis$^{1,2}$}
\author{A. I. Sokolov$^{1}$}
\email{ais2002@mail.ru}
\address{$^{1}$Division of Quantum Mechanics, Saint Petersburg State University,
Ulyanovskaya 1, Petergof, Saint Petersburg, 198504 Russia, \\
$^{2}$ITMO University, Kronverkskii ave 49, Saint Petersburg 197101, Russia}

\date{\today}

\begin{abstract}

{Critical fluctuations change the effective anisotropy of cubic
ferromagnet near the Curie point. If the crystal undergoes phase
transition into orthorhombic phase and the initial anisotropy is not too
strong, reduced anisotropy of nonlinear susceptibility acquires at $T_c$
the universal value $\delta_4^* = {{2v^*} \over {3(u^* + v^*)}}$ where
$u^*$ and $v^*$ -- coordinates of the cubic fixed point on the flow
diagram of renormalization group equations. In the paper, the critical
value of the reduced anisotropy is estimated within the pseudo-$\epsilon$
expansion approach. The six-loop pseudo-$\epsilon$ expansions for $u^*$,
$v^*$, and $\delta_4^*$ are derived for the arbitrary spin dimensionality
$n$. For cubic crystals ($n = 3$) higher-order coefficients of the
pseudo-$\epsilon$ expansions obtained turn out to be so small that use of
simple Pad\'e approximants yields reliable numerical results. Pad\'e
resummation of the pseudo-$\epsilon$ series for $u^*$, $v^*$, and
$\delta_4^*$ leads to the estimate $\delta_4^* = 0.079 \pm 0.006$
indicating that detection of the anisotropic critical behavior of cubic
ferromagnets in physical and computer experiments is certainly possible.}

\end{abstract}


\maketitle

\section{Introduction}

Thermal fluctuations of the order parameter are known to change effective
anisotropy of a cubic ferromagnet approaching the Curie point. As was
first discovered by Wilson and Fisher a system with cubic symmetry may
become isotropic under $T \to T_c$ provided its initial anisotropy is
small enough \cite {WF}. On the contrary, if the "bare" anisotropy is big
it increases further until the fluctuation-driven discontinuous phase
transition occurs \cite {W,KW,LP}. If a ferromagnet undergoes the
second-order phase transition, i. e. its initial anisotropy is not too
strong, and the low-temperature phase is orthorhombic, under $T \to T_c$
the anisotropy of magnetic subsystem acquires the universal value which
does not depend on its magnitude far from the Curie point. As was revealed
in the course of the study of critical behavior of $n$-vector cubic model,
the order parameter dimensionality $n$ plays a key role here: for $n <
n_c$ the system undergoing continuous phase transition demonstrates
isotropic critical behavior while for $n > n_c$ it remains anisotropic at
$T_c$ \cite{A}.

The numerical value of the marginal spin dimensionality $n_c$ separating
these two regimes is of prime physical importance since it determines the
true mode of the critical behavior of real cubic ferromagnets. Early
estimates of $n_c$ deduced from the lower-order renormalization-group (RG)
calculations \cite{KW,S77,NR} turned out to be in favor of the inequality
$n_c > 3$ implying that cubic ferromagnets should belong to the class of
universality of the three-dimensional (3D) Heisenberg model. The results
obtained within the high-temperature expansion approach seemed to support
this conclusion \cite{FVC}. However, the resummation of the three-loop RG
expansions \cite{MSI,Sh89} and subsequent higher-order RG analysis both in
three \cite{MSS,PS,CPV} and $(4 - \epsilon)$ \cite{KS} dimensions as well
as advanced lattice calculations shifted the value of $n_c$ downwards
fixing it below 3
\cite{MSI,Sh89,MSS,PS,CPV,KS,KT,KTS,SAS,FHY1,FHY2,V,PVr,HV}. Impressive
consensus between different field-theoretical approaches and resummation
techniques was achieved in the course of this study: 3D RG calculations
\cite{CPV}, resummed $\epsilon$ expansions \cite{SAS}, biased $\epsilon$
expansion technique \cite{CPV}, and pseudo-$\epsilon$ expansion analysis
\cite{FHY2} yielded $n_c = 2.89$, $n_c = 2.855$, $n_c = 2.87$, and $n_c =
2.862$, respectively. Thus, cubic ferromagnets undergoing second-order
phase transitions should demonstrate the anisotropic critical behavior
with a specific set of critical exponents.

On the other hand, since $n_c$ is very close to the physical value $n = 3$,
the cubic fixed point lies very near the Heisenberg fixed point at the RG
flow diagram. It implies that the critical exponents for both fixed points
should almost coincide. Indeed, the susceptibility exponent $\gamma$, for
instance, is equal to 1.3895(50) for the Heisenberg fixed point
\cite{GZ98} and to 1.390(6) for the cubic one \cite{CPV}. As a result,
measuring critical exponents in physical or computer experiments one can
not distinguish between the cubic and Heisenberg critical behaviors.

To clear up how the system with cubic symmetry behaves near $T_c$ one
should explore some alternative physical quantities. It was suggested
\cite{PS01} that nonlinear susceptibilities of different orders can play a
role of indicators of anisotropic critical behavior. The RG analysis of the
lowest-order nonlinear susceptibility $\chi^{(4)}$, in particular, has shown
that its reduced cubic anisotropy may be as large as 5$\%$ at criticality
\cite{PS01}, i. e. is certainly detectable experimentally. At the same time,
although this estimate was extracted from the longest -- six-loop -- RG
expansions available it is hardly believed to be quite reliable. The point
is that the anisotropy of $\chi^{(4)}$ is evaluated via the cubic fixed
point coordinates $u_{4}^*$ and $v_{4}^*$ and one of them ($v_{4}^*$)
being numerically small can not be found with high accuracy on the base
of the diverging series for $\beta$ functions.

In such a situation it looks natural to analyze the nonlinear
susceptibility of cubic ferromagnets and its anisotropy near $T_c$ using
alternative approaches. In this paper, the anisotropy of $\chi^{(4)}$ of
the 3D cubic model will be studied in the framework of the
pseudo-$\epsilon$ expansion technique. This approach invented by B. Nickel
(see Ref. 19 in the paper of Le Guillou and Zinn-Justin \cite{LGZ80})
exploits the idea that the fixed point location in three dimensions may be
found iteratively by means of introducing fictitious small parameter
$\tau$ into linear terms of the perturbative series for $\beta$ functions.
The method of pseudo-$\epsilon$ expansion proved to be very efficient when
used to estimate critical exponents and other universal quantities of
various three-dimensional systems
\cite{LGZ80,GZ98,FH97,FH99,FHY2,HDY01,DHY02,CP04,HID04,DHY04,NS14,SN14,NS15,SN16}.
Moreover, even in two dimensions where RG series are shorter and more
strongly divergent it enables one to get fair numerical results
\cite{LGZ80,COPS04,CP05,S2005,S13,NS13}. The numerical power of the
pseudo-$\epsilon$ expansion machinery, i. e. its ability to accelerate RG
iterations and to smooth oscillations of numerical estimates as functions
of the order of approximation, is so high that in many cases
pseudo-$\epsilon$ expansions ($\tau$-series) employed do not require
advanced resummation procedures. As a rule, use of Pad\'e approximants or
even direct summation turns out to be sufficient to lead to proper
numerical results.

The paper is organized as follows. In Section 2 the nonlinear
susceptibility $\chi^{(4)}$ of the cubic model is discussed, the parameter
$\delta_4$ characterizing its anisotropy is introduced, and the relation
between $\delta_4$ and effective quartic coupling constants $u_{4}$,
$v_{4}$ is presented. In Section 3 pseudo-$\epsilon$ expansions for the
cubic fixed point coordinates $u_{4}^*$ and $v_{4}^*$ and for the critical
value of the reduced anisotropy $\delta_4$ are derived for arbitrary $n$.
The $\tau$-series for $u_{4}^*$, $v_{4}^*$, and $\delta_4^*$ at $n = 3$
which are of particular physical interest are presented in Section 4. Here
the resummation of the pseudo-$\epsilon$ expansions obtained is carried
out and the numerical estimate for the reduced anisotropy at criticality
is presented. In Section 5 this estimate is compared with those given by
the 3D RG analysis, their deviation from each other is discussed, and the
robustness of the pseudo-$\epsilon$ expansion estimate is demonstrated.
Conclusion contains the summary of the main results obtained.

\section{Quartic coupling constants, nonlinear susceptibility, and its
anisotropy}
\label{sec:2}

Near Curie point the free energy of a cubic ferromagnet may be written down
as a series in powers of magnetization components $M_{\alpha}$:
\begin{eqnarray}
F(M_{\alpha}, m) = F(0, m) +  {\frac{1}{2}} m^{2 - \eta}
M_{\alpha}^2 + m^{1 - 2\eta}(u_4 + v_4 \delta_{\alpha \beta})
M_{\alpha}^2 M_{\beta}^2 + ...
\label{eq:1}
\end{eqnarray}
where $\eta$ is a Fisher exponent, $m$ being an inverse correlation length
while $u_4$ and $v_4$ are dimensionless effective coupling constants that
acquire under $T \to T_c$ the universal values. These coupling constants are
related to the fourth-order nonlinear susceptibility defined in a
conventional way:
\begin{equation}
\chi_{\alpha \beta \gamma \delta}^{(4)} =
{\frac{\partial^3 M_{\alpha}}{{\partial H_{\beta}}{\partial H_{\gamma}}
{\partial H_{\delta}}}} \Bigg\arrowvert_{H = 0}.
\label{eq:2} \\
\end{equation}

Of particular importance are the values of nonlinear susceptibility in
two cases, namely when i) an external field is parallel to a cubic axis
($\chi_{c}^{(4)}$) and ii) it is oriented along a unit cell space diagonal
($\chi_{d}^{(4)}$). For these two directions the difference between
corresponding values of nonlinear susceptibility turns out to be
maximal, i. e. the anisotropy is most pronounced. As may be readily
shown
\begin{equation}
\chi_{c}^{(4)} = - 24 {\frac{\chi^2}{m^3}} (u_4 + v_4),
\qquad \quad
\chi_{d}^{(4)} = - 24 {\frac{\chi^2}{m^3}} \Biggl(u_4 +
{\frac{v_4}{3}} \Biggr),
\label{eq:3} \\
\end{equation}
where $\chi$ is a linear susceptibility. To characterize the anisotropy
strength, we define a reduced parameter
\begin{equation}
\delta_4 = {\frac{\chi_{c}^{(4)} - \chi_{d}^{(4)}}{\chi_{c}^{(4)}}},
\label{eq:4}
\end{equation}
that is to be estimated at criticality.

Before starting our pseudo-$\epsilon$ expansion calculations let us remind
the estimate for the critical value of $\delta_4$ originating from the 3D
RG analysis. Coordinates of the cubic fixed point $u_4^*$ and $v_4^*$ in
three dimensions are known today from the higher-order RG calculations,
with resummed four-, five-, and six-loop RG expansions yielding close
numerical results \cite{MSS,CPV}. Considering the six-loop RG estimates as
the most reliable ones, we accept $u_4^* = 0.755 \pm 0.010$, $v_4^* =
0.067 \pm 0.014$ \cite{CPV}. Substitution of these numbers into Eqs. (3),
(4) gives
\begin{equation}
\delta_4^* = 0.054 \pm 0.012.
\label{eq:5} \\
\end{equation}
This value will be compared with those resulting from the
pseudo-$\epsilon$ expansions for universal values of the quartic couplings
as well as from the $\tau$-series for $\delta_4^*$ itself. These series
are derived in the next section.

\section{Pseudo-$\epsilon$ expansions for quartic couplings and reduced
anisotropy}
\label{sec:3}

Let us take as a starting point well-known Landau--Wilson Hamiltonian of
the 3D $n$-vector cubic model:
\begin{equation}
H = {\frac{1}{2}}
\int d^3x \Biggl[ m_0^2 \varphi_{\alpha}^2 +
(\nabla \varphi_{\alpha})^2
+ {\frac{u_0}{12}} \varphi_{\alpha}^2 \varphi_{\beta}^2
+ {\frac{v_0}{12}} \varphi_{\alpha}^4 \Biggr],
\label{eq:6}
\end{equation}
where $m_o^2$ is proportional to the deviation from the mean-field
transition temperature. Our analysis is based on the $\beta$ functions
(Gell-Mann--Low functions) of this model calculated within a massive
theory under zero-momenta normalizing conditions for the renormalized
Green function $G_R(p,m)$ and the four-point vertices $U_R({\bf
p_1,p_2,p_3},m,u,v)$, $V_R({\bf p_1,p_2,p_3},m,u,v)$:
\begin{eqnarray}
G_R^{-1}(0,m) = m^2, \ \ \ \ \
{\frac{\partial G_R^{-1}(p,m)}{\partial
p^2}} \Big\arrowvert_{p^2 = 0} = 1, \ \ \ \
\nonumber \\
U_R(0,0,0,m,u,v) = m u, \ \ \ \ \ V_R(0,0,0,m,u,v) = m v. \ \label{eq:7}
\end{eqnarray}
The value of the one-loop graph including the factor $(n + 8)$ is absorbed
in $u$ and $v$ in order to make the coefficient for $u^2$ term in the
$\beta_u$ function equal to unity. Quartic effective couplings $u$ and $v$
thus defined are related to $u_4$ and $v_4$ entering Eqs. (1), (3) in a
following manner:
\begin{eqnarray}
u = {\frac{n + 8}{2 \pi}} u_4, \qquad v = {\frac{n + 8}{2 \pi}} v_4.
\label{eq:8}
\end{eqnarray}

The perturbative expansions for $\beta$ functions of the model (6) are
known today up to the six-loop terms \cite{CPV}. Although the cubic fixed
point of the RG equations is stable only for $n > n_c$ we derive the
pseudo-$\epsilon$ expansions for coordinates of this point $u^*$, $v^*$
under arbitrary $n$. To find them the linear terms in $\beta$ functions
are replaced with $\tau u$ and $\tau v$, respectively, where $\tau$ is a
fictitious small parameter, and then the equations
\begin{equation}
\beta_u (u, v) = 0, \qquad \beta_v (u, v) = 0
\label{eq:9}
\end{equation}
are solved iteratively in $\tau$. This procedure results in the
pseudo-$\epsilon$ expansions ($\tau$-series) for $u^*$ and $v^*$:
\begin{eqnarray}
u^* &=& \frac{(n + 8)\tau}{n}\biggl[\frac{1}{3} - \frac{4(n - 1)(85 n -
478)}{2187}\frac{\tau}{n^2} + (4.5859515 - 8.6792769 n
\nonumber \\
&+& 5.4652167 n^2 - 1.1411824 n^3 + 0.0045537001 n^4)\frac{\tau^2}{n^4}
\nonumber \\
&+& (- 30.06975 + 78.40189 n - 79.05230 n^2 + 37.61208 n^3
\nonumber \\
&-& 8.073476 n^4 + 0.6043141 n^5 - 0.02075372 n^6)\frac{\tau^3}{n^6}
\nonumber \\
&+& (220.8250 - 733.6003 n + 1007.970 n^2 - 732.1351 n^3 + 298.0330 n^4
\nonumber \\
&-& 66.24082 n^5 + 7.381100 n^6 - 0.4608437 n^7 + 0.02278282
n^8)\frac{\tau^4}{n^8}
\nonumber \\
&+& (-1737.52 + 7014.10 n - 12174.6 n^2 + 11821.6 n^3 - 7004.25 n^4 +
2595.17 n^5
\nonumber \\
&-& 596.900 n^6 + 84.4687 n^7 - 7.88176 n^8 + 0.605252 n^9 - 0.0302510
n^{10})\frac{\tau^5}{n^{10}}\biggr], \label{eq:10}
\end{eqnarray}

\begin{eqnarray}
v^* &=& \frac{(n + 8)\tau}{n}\biggl[\frac{n - 4}{9} + \frac{4(n - 1)(77
n^2 + 494 n - 1912 )}{6561}\frac{\tau}{n^2} + (-6.1146020
\nonumber \\
&+& 12.064863 n - 7.8052925 n^2 + 1.5927396 n^3 + 0.026369322 n^4 +
0.00065967848 n^5)\frac{\tau^2}{n^4}
\nonumber \\
&+& (40.09300 - 107.7651 n + 111.0051 n^2 - 52.82124 n^3 + 10.49491 n^4
\nonumber \\
&-& 0.3739251 n^5 - 0.03604614 n^6 + 0.001331510 n^7)\frac{\tau^3}{n^6}
\nonumber \\
&+& (- 294.4333 + 1001.849 n - 1401.721 n^2 + 1025.114 n^3 - 408.9223 n^4
+ 82.35319 n^5
\nonumber \\
&-& 5.903868 n^6 - 0.1481744 n^7 + 0.02213800 n^8 - 0.004581224
n^9)\frac{\tau^4}{n^8}
\nonumber \\
&+& (2316.69 - 9538.73 n + 16819.2 n^2 - 16475.7 n^3 + 9715.12 n^4 -
3480.03 n^5 + 722.129 n^6
\nonumber \\
&-& 76.2769 n^7 + 2.81068 n^8 + 0.105962 n^9 - 0.0352014 n^{10} +
0.00445940 n^{11})\frac{\tau^5}{n^{10}}\biggr]. \label{eq:11}
\end{eqnarray}

These pseudo-$\epsilon$ expansions should obey some exact relations
appropriate to the systems with cubic anisotropy. For $n = 2$ the cubic
model is known to possess a specific symmetry: if the field
$\varphi_{\alpha}$ undergoes the transformation
\begin{equation}
\varphi_1 \to {\frac{\varphi_1 + \varphi_2}{\sqrt 2}}, \quad \varphi_2 \to
{\frac{\varphi_1 - \varphi_2}{\sqrt 2}},
\label{eq:12}
\end{equation}
quartic coupling constants are also transformed:
\begin{equation}
u \to u + {\frac{3}{2}} v, \quad v \to -v,
\label{eq:13}
\end{equation}
but the structure of the Hamiltonian itself remains unchanged \cite{WF}.
Since the RG functions of the problem are completely determined by the
structure of the Hamiltonian, the RG equations should be invariant with
respect to any transformation conserving this structure \cite{K}. This
should be also true for all the expressions relating various physical
quantities to each other. If, for example, one applies under $n = 2$ the
transformation (12) to the magnetization components in Eq. (1), the
free-energy expansion remains the same, provided $u_4$, $v_4$ are replaced
according to Eq. (13).

Since the transformation (12), (13) does not influence the structure of
the RG equations it should map corresponding flow diagram into itself.
This implies, in particular, that the locations of different fixed points
at the ($u$, $v$) plane should be related to each other. Indeed, if we
apply the transformation (13) to the Ising fixed point (0, $g_I^*$) lying
on the $v$ axis it will turn into the fixed point with coordinates
($3g_I^*/2$, $-g_I^*$), i. e. into the cubic fixed point, and vice versa.
Both fixed points have similar character of stability -- they are unstable
(saddle) fixed point. This similarity is quite natural since because of
the specific symmetry mentioned these points are equivalent being, in
fact, the same fixed point.

For $n = 2$ the pseudo-$\epsilon$ expansions for the coordinates of the
cubic fixed point resulting from (10), (11) are as follows
\begin{eqnarray}
u^* = \frac{5}{3}\tau + \frac{1540}{2187}\tau^2 + 0.0098952\tau^3 +
0.019972\tau^4 - 0.06872\tau^5 + 0.0669\tau^6, \label{eq:14}
\end{eqnarray}
\begin{eqnarray}
v^* = -\frac{10}{9}\tau - \frac{3080}{6561}\tau^2 - 0.0065968\tau^3 -
0.013315\tau^4 + 0.04581\tau^5 - 0.0446\tau^6. \label{eq:15}
\end{eqnarray}
Under the transformation (13) the cubic fixed point turns into the Ising
one. Hence, the sum $u^* + 3v^*/2$ should be equal to zero. As is seen,
$\tau$-series (14) and (15) do really obey the relation $u^* + 3v^*/2 =
0$. Moreover, due to the same symmetry the series (15) should coincide
with the pseudo-$\epsilon$ expansion for the Wilson fixed point coordinate
$g_I^*$ of the 3D Ising model. One can easily check, making obvious
rescaling, that the expansion (15) is identical to the six-loop
$\tau$-series for $g_I^*$ reported in Ref. 26.

At the end of this section we present the pseudo-$\epsilon$ expansion for
the reduced anisotropy $\delta_4^*$:
\begin{eqnarray}
\delta_4^*&=&\frac{2(n-4)}{3(n-1)}+\frac{16(n-2)}{9n(n-1)}\tau
+\frac{(n-2)}{n^3(n-1)}(-4.6627082 + 5.0533359 n - 0.76655300 n^2)\tau^2
\nonumber \\
&+&\frac{(n-2)}{n^5(n-1)}(24.45841 - 43.98925 n + 28.53713 n^2 - 5.962430
n^3 + 0.4618007 n^4)\tau^3
\nonumber \\
&+&\frac{(n-2)}{n^7(n-1)}(-160.375 + 403.068 n - 408.619 n^2 + 207.490 n^3
- 52.3182 n^4 + 6.04176 n^5 \nonumber \\ &-& 0.431021 n^6)\tau^4 +
\frac{(n-2)}{n^9(n-1)}(1177.7 - 3801.7 n + 5206.8 n^2 - 3920.4 n^3 +
1754.3 n^4
\nonumber \\
&-& 469.54 n^5 + 76.035 n^6 - 7.3777 n^7 + 0.52362 n^8)\tau^5.
\label{eq:16}
\end{eqnarray}
Note that all the terms of the series (16) apart from the first one vanish
when $n \to 2$. This reflects the specific symmetry discussed above which
enables, in particular, to find the exact value of the reduced anisotropy
under $n = 2$. Indeed, substituting the coordinates of the cubic fixed
point ($3g_I^*/2$, $-g_I^*$) into the relation
\begin{equation}
\delta_4^* = {{2v^*} \over {3(u^* + v^*)}} \label{eq:17}
\end{equation}
resulting from (3) and (4) we obtain $\delta_4^* = -4/3$ for $n = 2$. The
pseudo-$\epsilon$ expansion (16) is seen to be in accord with this result.

\section{Numerical estimates for cubic ferromagnets ($n = 3$)}
\label{sec:4}

Let us use the pseudo-$\epsilon$ expansions just obtained to estimate
the critical values of effective coupling constants and reduced
anisotropy for cubic ferromagnets, i. e. in the physical case $n = 3$.
Relevant $\tau$-series are as follows:
\begin{eqnarray}
u^* = \frac{11}{9} \tau + \frac{19624}{59049} \tau^2 - 0.12258455 \tau^3 -
0.0655945 \tau^4 - 0.061083 \tau^5 + 0.01269 \tau^6, \label{eq:18}
\end{eqnarray}
\begin{eqnarray}
v^* = -\frac{11}{27} \tau + \frac{23144}{177147} \tau^2 + 0.23233729
\tau^3 + 0.1283989 \tau^4 + 0.050252 \tau^5 + 0.02224 \tau^6,
\label{eq:19}
\end{eqnarray}
\begin{eqnarray}
\delta_4^* = -\frac{1}{3} + \frac{8}{27} \tau + 0.06663560 \tau^2 +
0.0529734 \tau^3 - 0.025228 \tau^4 + 0.03848 \tau^5. \label{eq:20}
\end{eqnarray}
These expansions originating from the diverging RG series for $\beta$
functions are divergent as well. Hence, to obtain from (18)--(20) the
numbers of interest proper resummation procedures should be applied. On
the other hand, since these series have small higher-order coefficients a
direct summation may result in fair numerical estimates. When applied to
the series (20) it yields, under $\tau = 1$, $\delta_4^* = 0.096$. This
number differs considerably from the estimate $\delta_4^* = 0.054$ given
by the 3D RG analysis \cite{CPV,PS01}. The estimate just found may be
corrected if we accept that the pseudo-$\epsilon$ expansions in hand are
asymptotic and one can get the best numerical results cutting off the
series by smallest terms. Such an optimally truncated direct summation of
$\tau$-series (20) leads to $\delta_4^* = 0.057$. An alternative way to
estimate $\delta_4^*$ with a help of direct summation is to find the
universal values of quartic coupling constants and to use the relation
(17). Direct summation of the series (18), (19) under $\tau = 1$ gives
$u^* = 1.318$, $v^* = 0.156$ resulting in $\delta_4^* = 0.071$.

The estimates of $\delta_4^*$ thus obtained are appreciably scattered and
obviously need to be refined. Since the coefficients of all the
$\tau$-series employed rapidly diminish simple resummation procedures not
addressing the Borel transformation may be applied. Use of Pad\'e
approximants [L/M] looks quite reasonable in our case. First we perform
the Pad\'e resummation of the series for $\delta_4^*$ itself. Pad\'e
triangle for the pseudo-$\epsilon$ expansion (20) is presented in Table I.
As one can see, Pad\'e estimates converge to the value
\begin{equation}
\delta_4^* = 0.077. \label{eq:21}
\end{equation}

An alternative estimate of the reduced anisotropy is obtained via Pad\'e
resummation of the $\tau$-series for $u^*$ and $v^*$. Corresponding Pad\'e
triangles are shown in Tables II and III. The values of coupling constants
this resummation technique results in are seen to be:
\begin{equation}
u^* = 1.322, \qquad  v^* = 0.182. \label{eq:22}
\end{equation}
The first number almost coincides with its counterpart $u^* = 1.321$ found
by means of the conform-Borel analysis of the original six-loop expansions
for $\beta$ functions \cite{CPV} while the second one differs
substantially from its 3D RG analog $v^* = 0.117$. A substitution of the
numbers (22) into (20) gives
\begin{equation}
\delta_4^* = 0.081, \label{eq:23}
\end{equation}
the value which is close to the Pad\'e estimate (21). Averaging over these
two numbers we obtain
\begin{equation}
\delta_4^* = 0.079 \pm 0.006. \label{eq:24}
\end{equation}
This value may be referred to as a final estimate the pseudo-$\epsilon$
expansion machinery yields for the reduced anisotropy of cubic ferromagnet
at criticality. In order to make this estimate realistic (certainly
conservative) we accepted its inaccuracy to be three times bigger than the
difference between Pad\'e-approximant-based values (23) and (21).

\section{Critical anisotropy versus $n_c$. What are the true values of $n_c$ and
$\delta_4^*$?}
\label{sec:5}

So, the pseudo-$\epsilon$ expansion technique leads to the value of the
reduced anisotropy which almost 1.5 times greater than its RG analogue
$\delta_4^* = 0.054$. How can one explain such a significant difference of
the two field-theoretical estimates obtained within the highest-order
available -- six-loop -- approximation? The roots of such a discrepancy
may lie in some peculiarity of the $\tau$-series for $\delta_4^*$: the
numerical value of the anisotropy is much smaller than the coefficients of
the first terms of the series (19), and it is calculated as a small
difference of big numbers. The same is true, as seen from (18), for the
coupling constant $v^*$.

This smallness, in its turn, reflects the fact that the boundary
dimensionality of the order parameter $n_c$ is close to 3. If $n_c$ coincided
with the physical value of $n$, then the anisotropy parameter would be equal
to zero at the critical point. Since the difference $3 - n_c$ is numerically
small ($0.1 \div 0.15$), the values of $\delta_4^*$ and $v^*$ turn out to be
small as well. However, precisely for this difference various field-theoretical
schemes provide significantly different estimates. For example, the processing
of the six-loop 3D RG expansions for $\beta$ functions of the cubic model by
means of the conform-Borel technique leads to $3 - n_c = 0.11$ \cite{CPV},
while the values of this difference obtained by the resummation of the
pseudo-$\epsilon$ and $\epsilon$ expansions for $n_c$ equal 0.138 \cite{FHY2}
and 0.145 \cite{SAS}, respectively. Since the first of the above mentioned
numbers differs from the others by tenths of percents, it is not surprising
that a difference of values of $\delta_4^*$, obtained within the same iteration
schemes, turns out to be significant.

To make the situation more transparent we calculate the cubic fixed point
coordinate $v^*$, closely related to $\delta_4^*$, as a function of $n$.
The curves $v^*(n)$ obtained by means of Pad\'e resummation of the
pseudo-$\epsilon$ expansion (11) under $n$ varying between 2.80 and 3.00
are shown in Fig. 1. Since diagonal Pad\'e approximants are known to
possess the best approximating properties, this figure contains curves
given by the highest-order near-diagonal approximants [3/2], [2/3], and by
the diagonal one [2/2]. The value of $v^*$ the six-loop 3D RG analysis
yields is also shown along with the values of marginal spin dimensionality
$n_c$ extracted from the 6-loop 3D RG series \cite{CPV}, from the
$\epsilon$ expansion \cite{CPV,SAS}, and from the pseudo-$\epsilon$
expansion \cite{FHY2} for $n_c$. Comparing 3D RG and pseudo-$\epsilon$
expansions estimates one can clearly see that the closer $n_c$ is to 3,
the smaller $v^*$ and reduced anisotropy are at criticality.

Keeping in mind this point it is worthy to determine a real accuracy the
pseudo-$\epsilon$ expansion technique provides when used to estimate the
marginal spin dimensionality. Obviously, the accuracy of such an estimate
is limited by the accuracy achieved in the course of evaluating the cubic
fixed point coordinate and reduced anisotropy for $n$ under which $v^*(n)$
and $\delta^*(n)$ vanish. This accuracy, in its turn, may be characterized
by the sensitivity of the numerical results with respect to iteration
procedures employed. To get an idea about this sensitivity we calculate
$n_c$ by means of solving equations
\begin{equation}
v^*(n) = 0, \qquad \qquad \delta_4^*(n) = 0 \label{eq:25}
\end{equation}
with left-hand sides given by the series (11) and (16) respectively
resummed using various Pad\'e approximants. The results of these
calculations are presented in Tables 4 and 5.

Alternative estimates of $n_c$ thus obtained -- 2.860 and 2.853 -- are
remarkably close to each other and to the value $n_c = 2.862$ given by the
Pad\'e resummation of the pseudo-$\epsilon$ series for $n_c$ itself
\cite{FHY2}. Hence, the pseudo-$\epsilon$ expansion machinery may be
thought of as a self-consistent iteration scheme. Moreover, the
pseudo-$\epsilon$ expansion estimates agree well with the values 2.87 and
2.855 the resummation of the $\epsilon$ expansion for $n_c$ yields
\cite{CPV,SAS}. All this enables us to accept
\begin{equation}
n_c = 2.86 \pm 0.01
\label{eq:26}
\end{equation}
as a most reliable numerical value of the marginal spin dimensionality.
This number differs appreciably from the six-loop 3D RG estimate $n_c =
2.89 \pm 0.04$ \cite{CPV} although both estimates are not in conflict
since their uncertainty bars overlap. 3D RG estimates, however, look less
accurate because of their sensitivity to the order of approximation.
Indeed, 5-loop 3D RG estimate $n_c = 2.91$ \cite{PS,CPV} and its 6-loop
analog differ from each other much stronger than their pseudo-$\epsilon$
expansion counterparts do. To demonstrate this we address the
pseudo-$\epsilon$ series for $n_c$
\begin{eqnarray}
n_c = 4 - \frac{4}{3} \tau + 0.2904231 \tau^2 -0.1896725 \tau^3
+0.1995126 \tau^4 -0.224651 \tau^5. \label{eq:27}
\end{eqnarray}
obtained from (11) by means of solving equation $v^*(n,\tau) = 0$
iteratively. Numerical values of $n_c$ the Pad\'e resummation of the
series (27) yields are shown in Table 6. As is seen, the pseudo-$\epsilon$
expansion estimates rapidly converge to the asymptotic value and the
deviation of the six-loop result from the five-loop one is really tiny
(smaller than 0.0001) \cite{U00}.

With the value of $n_c$ in hand, we can get an alternative estimate of
$v^*$. Since $v^*$ at $n = n_c$ is known to be zero and the difference $3
- n_c = 0.14$ is numerically small the value of $v^*$ for $n = 3$ may be
found from the power series
\begin{eqnarray}
v^* = \frac{d v^*}{d n}\Bigg\arrowvert_{n_c}(3 - n_c) +
\frac{1}{2}\frac{d^2 v^*}{d n^2}\Bigg\arrowvert_{n_c}(3 - n_c)^2 + ... .
\label{eq:28}
\end{eqnarray}
The derivatives entering (28) may be evaluated by means of the processing
of their pseudo-$\epsilon$ expansions that are easily extracted from the
series (11). Under $n = n_c = 2.86$ they are as follows:
\begin{eqnarray}
\frac{d v^*}{d n}\Bigg\arrowvert_{n_c} = 0.54580 \tau + 0.46307 \tau^2 +
0.17409 \tau^3 + 0.07830 \tau^4 - 0.01880 \tau^5 + 0.01746 \tau^6,
\label{eq:29}
\end{eqnarray}
\begin{eqnarray}
\frac{d^2 v^*}{d n^2}\Bigg\arrowvert_{n_c} = - 0.30398 \tau - 0.31453
\tau^2 - 0.19609 \tau^3 - 0.17131 \tau^4 - 0.07026 \tau^5 - 0.09451
\tau^6. \label{eq:30}
\end{eqnarray}
Good approximating properties of the $\tau$-series (29) are well pronounced,
the expansion (30) also looks suitable for resummation. Use of Pad\'e
approximants to process these series results in
\begin{eqnarray}
\frac{d v^*}{d n}\Bigg\arrowvert_{n_c} = 1.25, \qquad \quad \frac{d^2
v^*}{d n^2}\Bigg\arrowvert_{n_c} = -1.26.
\label{eq:31}
\end{eqnarray}
Taking into account that the higher-order contributions in (28) are
negligibly small and substituting numbers (31) into this power series we
get
\begin{eqnarray}
v^* = 0.163, \qquad \quad \delta_4^* = 0.0731. \label{eq:32}
\end{eqnarray}
The first number obtained differs from its counterpart (22) by 10\% only
while the second one is in agreement with the estimate (24). This allows
us to adopt that the true values of universal quantities $n_c$ and
$\delta_4^*$ lye within the uncertainty bars of our final estimates (24)
and (26). This confirms also the conclusion that the pseudo-$\epsilon$
expansion machinery is the self-consistent procedure powerful enough to
yield accurate numerical results.

So, use of the pseudo-$\epsilon$ expansion technique enabled us to refine
the value of reduced anisotropy $\delta_4^*$ which turned out to be 1.5
times higher than that given by the conventional 3D RG analysis. This
makes the arguments in favor of the possibility of experimental detection
of the anisotropic critical behavior in cubic ferromagnets \cite{PS01}
more convincing.

\section{Conclusion}

To summarize, we have analyzed the behavior of nonlinear susceptibility of
a cubic ferromagnet near Curie point within the pseudo-$\epsilon$
expansion approach. For the $n$-vector cubic model the six-loop
pseudo-$\epsilon$ expansions for the cubic fixed point coordinates and
reduced anisotropy have been derived for general $n$. Under the physical
value $n = 3$ these expansions have been found to possess a structure
favorable for getting numerical estimates. Having processed the
$\tau$-series for $u*$ and $v*$ and the pseudo-$\epsilon$ expansion for
$\delta_4^*$ by means of the Pad\'e approximants we've found the estimate
$\delta_4^* = 0.079 \pm 0.006$ which turned out to be considerably greater
than its 3D RG counterpart. This discrepancy has been argued to reflect
the fact that the value of the marginal spin dimensionality $n_c$ given by
the six-loop 3D RG analysis differs appreciably from that obtained within
the pseudo-$\epsilon$ expansion and $\epsilon$ expansion approaches. The
evaluation of $v^*$ and $\delta_4^*$ via $n_c$ has been performed and has
shown that alternative pseudo-$\epsilon$ expansion estimates are mutually
consistent what allows to consider them as certainly reliable. The
obtained value of $\delta_4^*$ is big enough to imply that the anisotropic
critical behavior of cubic ferromagnets predicted by the theory is
detectable in current physical and computer experiments.

\section*{ACKNOWLEDGMENT}

We gratefully acknowledge the support of Saint Petersburg State University
via Grant 11.38.636.2013 and of the Russian Foundation for Basic Research
under Project 15-02-04687.

\newpage

\begin{table}[t]
\caption{Pad\'e triangle for the pseudo-$\epsilon$ expansion (20) of the
reduced anisotropy $\delta_4$ in the Curie point. Approximant [2/1] has a
pole close to 1 ("dangerous"), corresponding estimate is not reliable and
therefore is given in brackets. The bottom line (RoC) shows the rate and
the character of convergence of Pad\'e estimates to the asymptotic value.
Here Pad\'e estimate of $k$-th order is the number given by corresponding
diagonal approximant [L/L] or by a half of the sum of the values given by
approximants [L/L$-$1] and [L$-$1/L] when a diagonal approximant is
absent.} \label{tab1}
\renewcommand{\tabcolsep}{0.4cm}
\begin{tabular}{{c}|*{6}{c}}
$M \setminus L$ & 0 & 1 & 2 & 3 & 4 & 5 \\
\hline
0 & $-$0.33333 & $-$0.03704  &   0.02960  & 0.08257   &  0.05734  & 0.09583 \\
1 & $-$0.17647 &    0.04893  &  (0.28797) & 0.06548   &  0.07258  & \\
2 & $-$0.11578 &    0.13681 &   0.07997  & 0.07700   & & \\
3 & $-$0.08139 &    0.03850  &  0.07688  & & & \\
4 & $-$0.06127 &    0.17108  & & & & \\
5 & $-$0.04705 & & & & & \\
\hline RoC & $-$0.33333 & $-$0.10675 & 0.04893  & (0.21239) & 0.07997 &
0.07694
\end{tabular}
\end{table}

\begin{table}[t]
\caption{Pad\'e table for the pseudo-$\epsilon$ expansion (18) of the
coupling constant $u^*$. Approximants are constructed for $u^*/\tau$, i.
e. neglecting the insufficient factor $\tau$. Approximants [3/1], [3/2]
and [2/2] have dangerous poles, therefore corresponding estimates are not
reliable; in the table they are bracketed. The bottom line (RoC) indicates
the character of convergence of Pad\'e estimates to the asymptotic value.
Here the Pad\'e estimate of $k$-th order is the number obtained in the
same manner as in the case of $\delta_4^*$ (Table I). The final value $u^*
= 1.3218$ is the result of averaging over three working approximants
[4/1], [2/3], and [1/4].} \label{tab2}
\renewcommand{\tabcolsep}{0.4cm}
\begin{tabular}{{c}|*{6}{c}}
$M \setminus L$ & 0 & 1 & 2 & 3 & 4 & 5 \\
\hline
0 & 1.2222  & 1.5546  & 1.4320    & 1.3664    & 1.3053   & 1.3180 \\
1 & 1.6787  & 1.4650  & 1.2909    & (0.4782)  & 1.3158   & \\
2 & 1.3545  & 1.3832  & (0.9025)  & (1.2483)  & & \\
3 & 1.3868  & 1.3629  & 1.3197  & & & \\
4 & 1.3004  & 1.3300  & & & & \\
5 & 1.3472  & & & & & \\
\hline RoC  & 1.2222  & 1.6166  & 1.4650  & 1.3371  & (0.9025) & 1.3218
\end{tabular}
\end{table}

\begin{table}[t]
\caption{Pad\'e triangle for pseudo-$\epsilon$ expansion (19) of the
coupling constant $v^*$. Approximants are constructed for $v^*/\tau$, i.
e. with the factor $\tau$ omitted. Approximant [1/1] has a dangerous pole,
corresponding estimate is unreliable and presented in brackets. The
convergence of Pad\'e estimates to the asymptotic value is illustrated by
the bottom line (RoC). Here the Pad\'e estimate of $k$-th order is the
same as in Table 1.} \label{tab3}
\renewcommand{\tabcolsep}{0.4cm}
\begin{tabular}{{c}|*{6}{c}}
$M \setminus L$ & 0 & 1 & 2 & 3 & 4 & 5 \\
\hline
0 & $-$0.4074  &  $-$0.2768   & $-$0.0444  & 0.0840  & 0.1342  & 0.1565 \\
1 & $-$0.3085  & ($-$0.5753)  & 0.2426     & 0.1665  & 0.1741  & \\
2 & $-$0.2043  &    0.0416    & 0.1507     & 0.1729  & & \\
3 & $-$0.1505  &    0.1905    & 0.1906     & & & \\
4 & $-$0.1149  &    0.1906    & & & & \\
5 & $-$0.0900  & & & & & \\
\hline RoC & $-$0.4074 & $-$0.2926 & ($-$0.5753) & 0.1416 & 0.1507 &
0.1817
\end{tabular}
\end{table}

\begin{table}[t]
\caption{Values of $n_c$ calculated by solving the equation $v^*(n) = 0$
with the left-hand side equal to Pad\'e-resummed pseudo-$\epsilon$
expansion (11). Pad\'e approximants [L/M] are constructed for $v^*/\tau$,
i. e. neglecting the trivial factor $\tau$. Approximants [1/1] and [1/3]
have poles very close to the knots of interest, corresponding estimates
are not reliable and absent in the table. The bottom line (RoC) indicates
the character of convergence of estimates to the asymptotic value. Here
the estimate of $k$-th order is the number obtained in the same manner as
in Table I.} \label{tab4}
\renewcommand{\tabcolsep}{0.4cm}
\begin{tabular}{{c}|*{6}{c}}
$M \setminus L$ & 0 & 1 & 2 & 3 & 4 & 5 \\
\hline
0 & 4.0000  & 3.3268  & 3.0419  & 2.9278  & 2.8842  & 2.8673 \\
1 & 4.0000  &    -    & 2.8056  & 2.8479  & 2.8571  & \\
2 & 4.0000  & 2.9651  & 2.8542  & 2.8601  &         & \\
3 & 4.0000  &    -    & 2.8594  & & & \\
4 & 4.0000  & 2.8723  & & & & \\
5 & 4.0000  & & & & & \\
\hline RoC  & 4.0000  & 3.6634  &    -    & 2.8854  & 2.8542 & 2.8598
\end{tabular}
\end{table}

\begin{table}[t]
\caption{Values of $n_c$ extracted from the equation $\delta^*(n) = 0$
with the left-hand side given by $\tau$-series (16) resummed using Pad\'e
approximants [L/M]. Two estimates are absent since approximant [2/1] is
spoiled with dangerous pole while approximant [1/4] has no real knots. The
bottom line (RoC) indicates the character of convergence of estimates to
the asymptotic value where the estimate of $k$-th order is the number
obtained in the same way as in Table I.} \label{tab5}
\renewcommand{\tabcolsep}{0.4cm}
\begin{tabular}{{c}|*{6}{c}}
$M \setminus L$ & 0 & 1 & 2 & 3 & 4 & 5 \\
\hline
0 & 4.0000  &  3.0704  & 2.9457   & 2.8517  & 2.8899  & 2.8217 \\
1 & 4.0000  &  2.9092  &    -     & 2.8784  & 2.8649  & \\
2 & 4.0000  &  2.7255  & 2.8501   & 2.8528  &         & \\
3 & 4.0000  &  2.9244  & 2.8528   &         &         & \\
4 & 4.0000  &     -    & & & & \\
5 & 4.0000  & & & & & \\
\hline RoC  & 4.0000  & 3.5352    & 2.9092  &    -    & 2.8501 & 2.8528
\end{tabular}
\end{table}

\begin{table}[t]
\caption{Pad\'e triangle for pseudo-$\epsilon$ expansion (27) of the
marginal spin dimensionality $n_c$. Approximant [1/2] has a pole very
close to 1, corresponding estimate being unreliable is bracketed. The
convergence of Pad\'e estimates to the asymptotic value is illustrated by
the bottom line (RoC) where the Pad\'e estimate of $k$-th order is the
number obtained in the same way as in Table 1.} \label{tab6}
\renewcommand{\tabcolsep}{0.4cm}
\begin{tabular}{{c}|*{6}{c}}
$M \setminus L$ & 0 & 1 & 2 & 3 & 4 & 5 \\
\hline
0  &  4.00000  &  2.66667  &  2.95709  &  2.76742  &  2.96693  &  2.74228  \\
1  &  3.00000  &  2.90515  &  2.84235  &  2.86456  &  2.86126  &    \\
2  &  2.91579  & (2.06451) &  2.86156  &  2.86162  &  &  \\
3  &  2.84113  &  2.87113  &  2.86162  &  &  &  \\
4  &  2.89218  &  2.86384  &  &  &  &  \\
5  &  2.82985  &  &  &  &  &  \\
\hline RoC  & 4.00000  & 2.83333    & 2.90915  &    -    & 2.86156 &
2.86162
\end{tabular}
\end{table}

\begin{figure}
\begin{center}
\includegraphics[width=\linewidth]{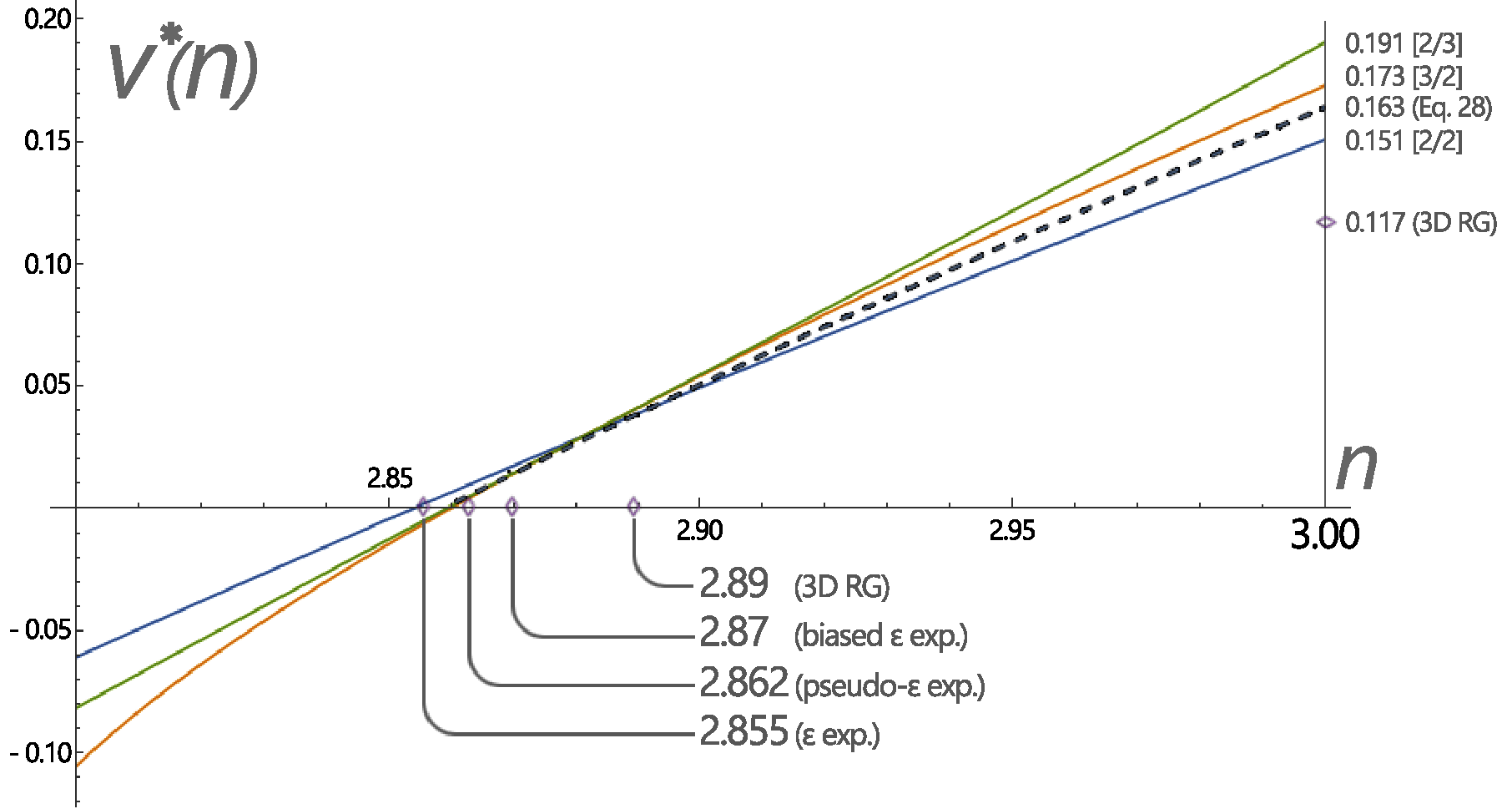}
\caption{(Color online) The cubic fixed point coordinate $v^*$ as a
function of $n$ obtained by means of Pad\'e resummation of $\tau$-series
(11). Three curves given by the diagonal [2/2] and near-diagonal
approximants [3/2], [2/3] for $v^*/\tau$, i. e. with insignificant factor
$\tau$ omitted, are shown. The values of marginal spin dimensionality
$n_c$ found by the analysis of 6-loop 3D RG series \cite{CPV}(2.89) and
extracted from the $\epsilon$ expansion \cite{CPV,SAS} (2.87, 2.855) and
from the pseudo-$\epsilon$ expansion \cite{FHY2} (2.862) for $n_c$ are
marked with diamonds. The diamond on the right vertical axis shows the
value of $v^*$ the six-loop 3D RG analysis yields. The dashed line is the
curve $v^*(n)$ given by the powers series in $(n - n_c)$ which arrives to
0.163 when the spin dimensionality approaches the physical value $n = 3$.}
\label{fig1}
\end{center}
\end{figure}


\begin{thebibliography}{999}
\bibitem{U00} The series (27) and numbers in Table 6 are in a very good aggrement
with those obtained in Ref. [19]. In this paper, the pseudo-$\epsilon$
expansion for $n_c$ was resummed by means of the Pad\'e-Borel-Leroy
technique and high stability of the numerical results with respect to the
resummation procedure was demonstrated.
\end{thebibliography}
\end{document}